\documentclass[a4paper]{article}
\usepackage{graphicx,subfigure}

\begin{document}

\title{Power of earthquake cluster detection tests}

\author{F. Dimer de Oliveira 
     \\ \small{\it Risk Frontiers,  Macquarie University}, 
     \\ \small{\it North Ryde, NSW, 2109, Australia.} \\ \small{felipe.dimer-de-oliveira@mq.edu.au} }

\maketitle

\begin{abstract}

Testing the global earthquake catalogue for indications of non-Poissonian attributes has been an area of intense research, especially since the 2011 Tohoku earthquake. The usual approach is to test statistically for the hypothesis that the global earthquake catalogue is well explained by a Poissonian process. 
%
%
%
In this paper we analyse one aspect of this problem which has been disregarded by the literature: the power of such tests to detect non-Poissonian features if they existed; that is, the probability of type II statistical errors. We argue that the low frequency of large events and the brevity of our earthquake catalogues reduces the power of the statistical tests so that an unequivocal answer for this question is not granted. We do this by providing a counter example of a stochastic process that is clustered by construction and by analysing the resulting distribution of p-values given by the current tests.


\end{abstract}

\section{Introduction}

Several investigators have proposed the presence of two temporal clusters of very large earthquakes during the past century, e.g. \cite{bufeandperkins2005,ammon2011}. The first cluster occurred in the middle of last century and included the 1952 Mw 9.0 Kamchatka earthquake, the 1960 Mw 9.5 Chile earthquake and the Mw 9.2 Alaska earthquake (\cite{bufeandperkins2005}). The second aparent cluster began with the occurrence of the Mw 9.15 Sumatra earthquake of 26 December 2004 and has continued with the Mw 8.8 Chile earthquake on 27 February 2010 and the Mw 9.0 Tohoku earthquake on 11 March 2011 \cite{bufeandperkins2011,ammon2011}.This recent cluster has given rise to debate about whether the observed temporal clustering of these very large earthquakes has some physical cause or has occurred by random chance \cite{kerr2011}.  

\cite{michael2011} used three statistical tests to conclude that the global clustering can be explained by the random variability in a Poisson process.  His first test was an analysis of inter-event times using a one-sided Kolmogorov-Smirnov test. The second test showed that the occurrence of very large earthquakes is not correlated with the occurrence of smaller events. The third test demonstrated that temporal clustering in seismic moment release occurs in about $50\%$ of the samples when the number of events is drawn from a Poisson distribution and is not constrained as in the modeling of \cite{bufeandperkins2005,bufeandperkins2011}. In another article, \cite{shearerandstark2011} reach the same conclusions testing for the Poissonian hypothesis using a different set of statistical quantities.

The purpose of this paper is to discuss the power of traditional statistical tests to establish unequivocally the existence or not of earthquake clusters for catalogues with small numbers of events and not amenable to experimental repeatability. In general, to study the power of statistical tests we need to enunciate an alternative hypothesis and calculate the probability of correctly rejecting a false hypothesis \cite{merrilandfox1970}; this is not the case for most studies of earthquake clusters since, to our knowledge, no stochastic process other than Poisson has been widely hypothesized and tested for in the earthquake catalogue. The objective of our study is to determine the probability with which a random sample of a contrived non-Poissonian process is rejected in a test in which the null hypothesis is a Poisson process.

To aid the discussion we have devised a stochastic process which is clustered by construction and whose samples play the role of earthquake catalogues with a given magnitude threshold and de-clustered to remove aftershocks. To each one of these samples we apply a specific statistical test and use the set of p-values obtained in this way to calculate their probability distribution. This distribution will inform us the probability that any random sample of this process will pass or fail a test for Poissonian statistics.





To justify the merits of our analysis, we start by observing that Poisson is the unique discrete stochastic process that satisfies two conditions: lack of memory (Markov assumption) and a constant probability of event occurrence through time \cite{gardiner2003}. The exceptional character of this process makes it a valuable tool in the natural sciences since lack of memory and time independence can be inferred either \emph{a priori} or \emph{a posteriori} - in this case by showing that the observed data fits well a Poisson distribution. Statistical inference of this sort is usually obtained through consensus of a large number of independent experiments, sometimes aided by theoretical models \footnote{One example is photo counting experiments of stable laser light, in which a Poisson distribution is derived from first principles quantum mechanics and experimentally verified to an large degree of confidence ($ \ll 1\% $)}

On the other hand, there is an infinite number of processes not satisfying one or both conditions. This becomes relevant when the available data is limited and a consensus view cannot be established since short data series could be explained adequately by more than one stochastic process. It is granted that, even in such occasions a Poisson distribution can be postulated on arguments of simplicity and plausibility, which, while scientific valid does not constitute objectively an explanation for a phenomenon. Otherwise, a Poisson process should be regarded as only one among many possible explanations. 

One obvious question that arises from these considerations regards how much data is enough so that an inference exercise can assert beyond reasonable doubt which model explains the data observed. The answer to this question lies in the scale of the stochastic process as compared with the length of the observed, which is illustrated by study.


\section{Description of the process}

The stochastic process that we use to assess the skill of Poissonian properties tests was devised as a theoretical artefact and not as a statistical model for the earthquake catalogue or associated with any particular physical reality. It was designed to convey in a synthetic manner the features of clustered data in which clusters may occur randomly and at relative low frequencies. 


This process is constructed by generating a Poisson series at low event rates (from 2 to 3 cluster per century)  equivalent to 110 years of observations. A cluster is a period of increased rate of event occurrence; we express this by inserting in each cluster occurrence a Poisson sample with a 10-fold increase in frequency (3 to 4 events per decade) and a duration of 15 years.
%
%
%
Clusters are not allowed to overlap, but can neighbour each other to form mega-clusters of $\approx 30 $ years. Each particular choice of parameter will give rise to a particular distribution of the 110-years average event rate. We have chosen the parameters above to coincide with the general scale of the observed global earthquake catalogue: average event frequencies will range from 1 to 2 evens per decade, with a large variability for the averages derived from any single 110 years sample (standard deviation $\sigma\approx 0.32$). As a reference, the global earthquake catalogue for a cut-off magnitude of 8.3 is approximately 2 events per decade.

The samples generated by this process will produce clusters which are aperiodic and could be interpreted either as due to self sustained triggering (one event increases the probability of another) or as an overall increase in event rates due to a single underlying cause. In either case, the samples are, on average, clustered enough so that the the p-value distribution is that of a non-uniform distribution and skewed to the left. To best represent the variability in the genesis of earthquakes both clusters and events within clusters are subjected to full variability of Poisson process - this means a non-zero probability of entire centuries with no clusters. 

We do not consider the problem of cluster detection for catalogues where a true Poisson noise or another independent clustered process was added to the background, nor of a periodic cyclic process - we assume it to be self evident that this entails a greater similarity with a Poissonian process. The approach we take here is conservative insofar as the process we envisage produces samples which are more clustered than a true Poissonian process.

\section{Analysis of p-values}

\subsection{p-Values distribution}

P-values distribution were obtained by generating 10,000 independent samples of our process and by performing three different statistical tests in each independent sample to obtain the corresponding p-value. The tests we have chosen are three: (a) Kolmogorov-Smirnov test on inter-event time distribution (b) Pearson $\chi$-square test on event count and (c) same same test on multiple event inter-time distribution. 

Test (a) was performed under the null hypothesis that inter-event times follow an exponential distribution characteristic of a Poisson process 
\begin{equation}
P(t)=\exp(-\lambda t)
\end{equation}
where $\lambda$ is the average event rate per year and $t$ is the time measured in years. The annual event rate $\lambda$ is re-calculated for each sample, simulating our ignorance of the true event rate. Test (b) is the usual Pearson $\chi$-square that tests for similarities in the histogram distribution between the samples and a Poisson distribution. Test (c) performs the same test as (b) on the inter-event time distribution for multiple events using the corresponding Poisson distribution for null hypothesis. These computations were performed using Mathematica{\copyright} software package \cite{mathematica2011}. Our procedure was tested by performing these same tests with Poisson generated samples, which correctly output uniform distributions of p-values.

\subsection{results}

\begin{figure}[h]
 \noindent\includegraphics[width=20pc]{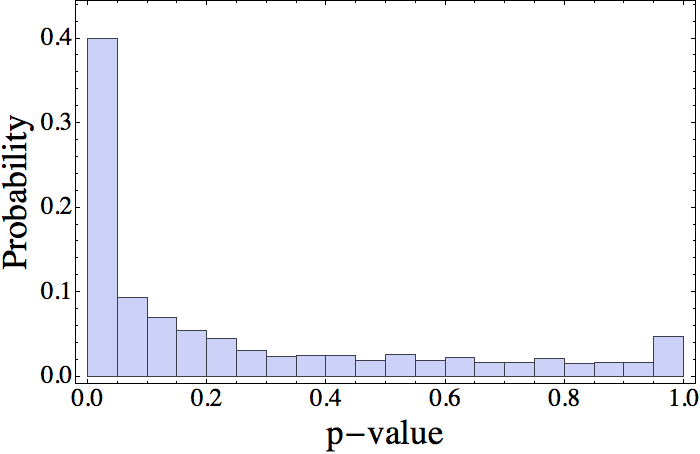}
 \caption{Distribution of p-values obtained by testing the fit of an exponential distribution with the Kolmogorov-Smirnov test for inter-event times in a clustered-by-construction process. The bins of this histogram have width of 5\%, thus the first bin corresponds to the probability of hypothesis rejection under the 5\% significance level.}
 \label{figure1}
 \end{figure}

A sample result is the p-value distribution shown in Figure \ref{figure1}. It represents the probability of measuring a given p-value for test (a) for a single 110 years clustered sample. We have adopted the most common view of rejecting a hypothesis for p-values smaller than 5\% -- a criterion that we return to in later discussions.
%
%
We have repeated the same process by varying the parameters of our process to estimate the power of detection of test (a) as the frequency of clusters and of events within clusters vary. The process parameters for the result of Figure \ref{figure1} are 3 clusters per century and 4 events per decade over a 15 year cluster period -- or 3-by-4 in short. With these parameters, the average annual frequency of events is 0.12 events/year, with a 70th and 90th percentiles above median of 0.15 and 0.2 events/year approximately. In the histogram of Figure \ref{figure1} bins are plotted in intervals of 5\% and we can see that the probability of a p-value smaller than $5\%$ is $\approx 40\%$. If the 5\% significance level is strictly adopted, this is the chance that an observer would correctly reject the hypothesis and implies a type-II error probability of 60\%.

\begin{figure}[h]
 \noindent\includegraphics[width=20pc]{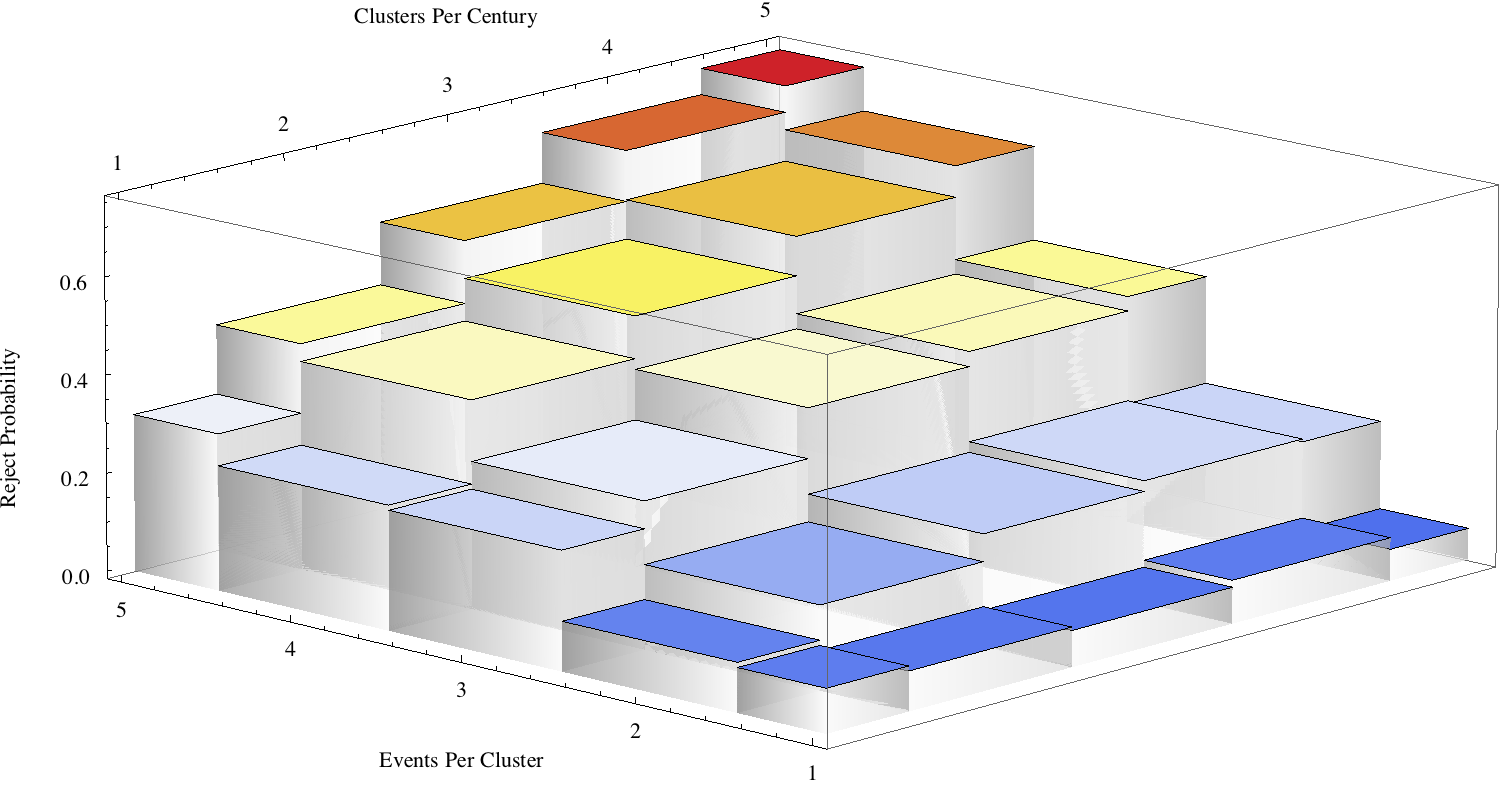}
 \caption{Probabilities of p-values less than 5\% by testing inter-event time distribution for a Poisson distribution with a Kolmogorov-Smirnov test.}
 \label{figure2}
 \end{figure}

We performed test (a) varying the parameters of the generating Poisson processes and show in Figure \ref{figure2} the probabilities of obtaining a p-value smaller than 5\%. As expected, for low frequency of events per cluster the process looks more like a Poisson process and is less frequently rejected; the samples of this process are maximally non-Poissonian for high cluster and event frequencies, with probability of a correct "reject" above 70\%. Average event frequencies for these values vary from 2 to 1 events per decade in the 5-by-5 
%
%
and 3-by-4 cases respectively (see Figure \ref{figure1}).
%

We will not discuss test (b), which has shown to be the less skilled of all, with probability of rejection on the order of 20\%. 

The most successful test among those we studied is (c). In it, we generated the null hypothesis distribution from the inter-$n$-event time distribution from samples of a Poisson process, against which the clustered samples were tested. In this test, the Poisson hypothesis distribution is assumed to have a \emph{known average event frequency} given by the long term mean over \emph{all samples}. 

\begin{figure}[h]

\begin{center}
 \subfigure[$\lambda = 0.12 $]{
  \label{figure3a} 
 \includegraphics[width=15pc]{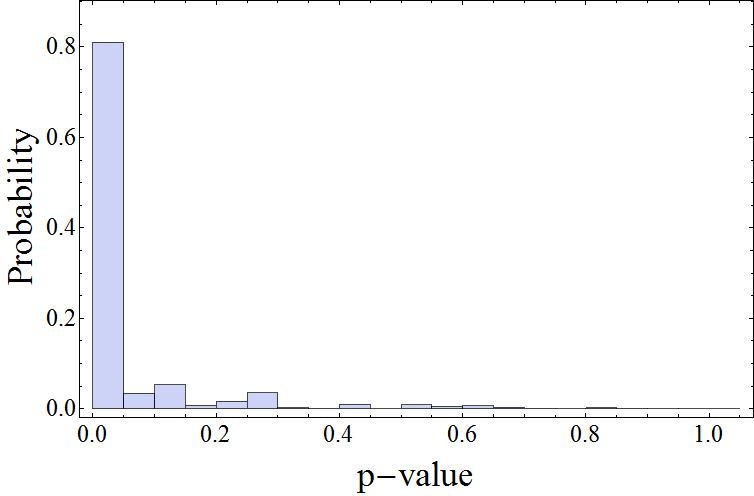} 
 }
  \subfigure[$\lambda = 0.15 $]{
  \label{figure3b} 
  \includegraphics[width=15pc]{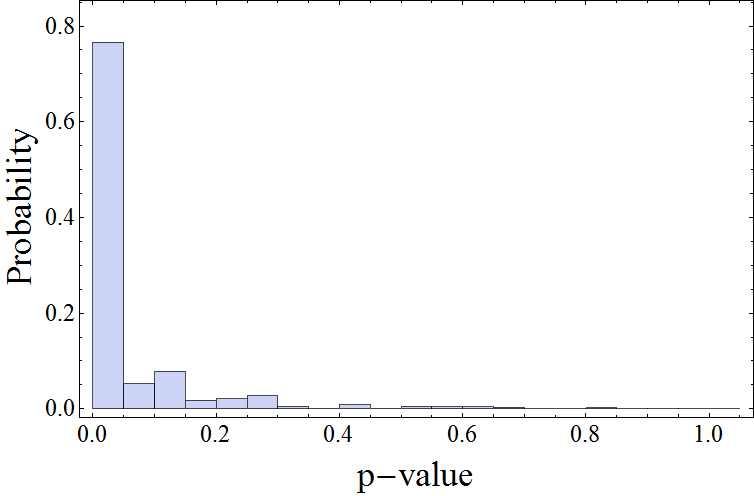}
  }
 \subfigure[$\lambda = 0.20 $]{
  \label{figure3c}   
   \includegraphics[width=15pc]{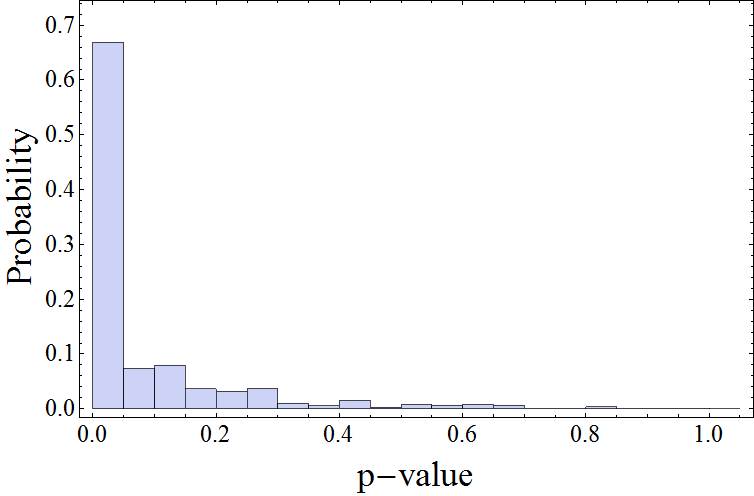}
} 
\end{center}
 \caption{Distributions of p-values using Pearson $\chi$-square for a Poisson hypothesis testing multiple-events inter event times. The stochastic process has parameters of 3 clusters per century and 4 events per decades within each cluster. The samples were tested for hypothesized Poisson distribution of different annual frequency parameters $\lambda$. The average over all 2000$\times$110 years samples is $\lambda=0.12$. The values $\lambda=0.15$ and $\lambda=0.20$ correspond approximately to the 70\% and 90\% quantiles of the distribution of average frequency for all samples respectively.}
 \label{figure3}
\end{figure}

A more sensible approach is to take into consideration the probability that a single 110 years average event rate will be that of the long term average. This can be done, brute force, by generating one null hypothesis for each sample to be tested against all samples,
%
%
%
%
thus accounting for the chance that a particular 110-years and the long term average event rate are the same. In the interest of focusing on the essential points, we show the distributions for this test for the cases where the Poisson frequency is the long term average (over all samples), the 70\% and 90\% quantiles above median for the same parameters as those shown in Figure \ref{figure1} (3 clusters per century 4 events per decade within clusters). The results are explained in Figures \ref{figure3} from (a) through (c). In it we see that, if we pick a sample whose value is the same as the average event rate, the test will detect correctly the non-Poissonian nature of this process 80\% of times. For samples whose average event rate is above the 90\% quantile, the probability of correctly rejecting the null hypothesis drops to about 70\%. The true value of the power for this test will depend on the degree of confidence in the estimation of the average event frequency and our belief of how accurately a single 110-years sample will inform on the "true" long term sample. This reflects the fact that this process unfolds on time-scales greater than 110 years.

\section{Discussion}

We stress that our presentation is not a claim that the stochastic process we devised intends to be a realistic model of the genesis of mega-earthquakes on a global scale. The results we have show are solely an illustration of the pitfalls of statistical tests and of type II errors. At the heart of these issues lies the statistical variability of the process we used, which can be plainly expressed by saying that some of the samples are more Poissonian than others. It is the assessment of differences \emph{between} trajectories enables us to determine the falsehood of the Poissonian hypothesis. It remains to be seen the results of a similar study using plausible non-Poissonian processes, and the effects of the introduction of a Poissonian background.

As we noted before, arguments of plausibility and simplicity based on Markovian and time-independence assumptions provide solid grounds for hypothesising a Poisson process as a likely candidate to explain the global earthquake catalogue. However, when viewed \emph{only on the merits of the observed data}, the probability of type II statistical error, such as those we computed here, must be taken into account. Our degree of belief in a given premise is explicitly manifest in Bayesian inference through the assignment of prior probabilities \cite{mackay2003}; any argument that deems to inquire the data \emph{alone} should state clearly its prior, whether equal probabilities or weighted towards a Poisson distribution. [Such argument can be made not only as a matter of scientific clarity but as as an aid to scientific imagination.] 

Another aspect we discuss here regards the levels of significance commonly used in statistics. We have used 5\% as the \emph{par excellence} standard in rejecting an hypothesis. Economics can provide the basis for a rational approach to choosing levels of significance by considering the costs of taking an erroneous decision based on a failed test \cite{merrilandfox1970}. 

Such type of argument in the case of earthquake clusters is not straightforward nor scientifically objective. Regarding the unequivocal establishment of a scientific statement, the setting of a level of significance should take into consideration the probabilities such as those we have derived here. In Figure \ref{figure1}, for example, the probability of a p-value above 20\% is non-trivial ($>10\%$). 

In light of this discussion, we can consider the recent results of earthquake cluster detection. Test (a) corresponds to the first test of \cite{michael2011} which was applied to the global catalogue at a large cut-off magnitude of $M_w=9$, which corresponds to a frequency of approximately 0.04 event/year, and he reports p-values as low as 0.12. We have shown that the same test would not be accurate event at much lower cut-off implying an event frequency of 0.1-0.2 event/year (corresponding magnitude thresholds between 8.4 and 8.3). From \cite{shearerandstark2011}, we are mostly interested in their multinomial test as it is equivalent to our case (c), which we assessed as the most powerful; in their work, the  p-values reported for a $M_w>8$ magnitude cut-off range between 35 to 25\% depending on the de-clustering undertaken. The average event rate for these magnitudes is well above any we have analysed here (0.8-0.7 event/year). More relevant to this discussion is \cite{shearerandstark2011} assertion that: "(...) the null hypothesis that times of large earthquakes follow a homogeneous Poisson process would not be rejected by any of the tests". Based on our discussion, the criteria to accept (or reject) a hypothesis is not a clear-cut line. These considerations go beyond the specifics of cluster detection (e.g. \cite{stumpf2012}

This work does not suggest that clustering is a real phenomenon -- considering that our test-process is highly contrived. This is a tentative way to introduce some objectivity into assertions such as "random variability explains earthquake catalogue": what would really be meant by "explains"? The fact that a given series of events has a "reasonable" probability according to such a process (and we have yet to define what we mean by reasonable) -- at most we could say it is consistent when favouring some prior, but as far a such limited data set is available such strong conclusions must not be taken for granted.

\section{acknowledgments}
The author acknowledges the contributions of Profs Paul Somerville, Rob Van den Honert and John McAneney to this paper, and the financial support of Lloyd’s of London.

\section{Figures}

\begin{thebibliography}{}


\bibitem[{\textit{Ammon et al.}(2011)}]{ammon2011}
Ammon, C. J., R. C. Aster, T. Lay and D. W. Simpson (2011), The Tohoku earthquake and a 110 year spatiotemporal record of global seismic strain release, 
{\it Seismological Society of America meeting, Memphis, April 14, 2011,}

\bibitem[{\textit{Bufe and Perkins.}(2005)}]{bufeandperkins2005}
Bufe, C. G. and D. M. Perkins (2005), Evidence for a global seismic moment release sequence, 
{\it Bull. Seismol. Soc. Am.,} \textit{95}, 833-–843

\bibitem[{\textit{Bufe and Perkins.}(2011)}]{bufeandperkins2011}
Bufe, C. G., and D. M. Perkins (2011), The 2011 Tohoku earthquake: Resumption of temporal clustering of Earth’s megaquakes, 
{\it Seismological Society of America meeting, Memphis, April 14, 2011,}

\bibitem[{\textit{Kerr}(2011)}]{kerr2011}
Kerr, R. (2011).  More earthquakes on the way? 
{\it Science,} \textit{332} 411 

\bibitem[{\textit{Michael}(2011)}]{michael2011}
Michael, A. (2011),  Random variability explains apparent global clustering of large earthquakes,
{\it Geo. Res. Lett.,} \textit{38}, L21301.

\bibitem[{\textit{Shearer and Stark}(2011)}]{shearerandstark2011} 
Shearer, P. M. and Stark P. B. (2011), Global risk of big earthquakes has not recently increased,
{\it PNAS }, published ahead of print December 19, 2011, doi:10.1073/pnas.1118525109

\bibitem[{\textit{Merril and Fox}(1970)}]{merrilandfox1970}
Merril W. C. and K. A. Fox, Introduction to Economic Statistics
{\it John Wiley}, 1970

\bibitem[{\textit{Gardiner}(2003)}]{gardiner2003}
Gardiner C. W., Handbook of stochastic methods
{\it Springer Verlag}, 2003

\bibitem[{\textit{Mackay}(2003)}]{mackay2003}
Mackay, D. J. C. (2003),  Information Theory, Inference, and Learning Algorithms,
{\it Cambridge Press}


\bibitem[{\textit{Mathematica 8}(2011)}]{mathematica2011}
Mathematica software documentation 
{\it http://reference.wolfram.com/mathematica/ref/PearsonChiSquareTest.html}

\bibitem[{\textit{Stumpf and Porter}(2012)}]{stumpf2012} 
Stumpf, Michael P. H. and Porter, Mason A. (2012), Critical Truths About Power Laws,
{\it Science }, \textit{335}, 665--666



\end{thebibliography}
\end{document}